# Chemical treatments of Monolayer Transition Metal Dichalcogenides and Their Prospect in Optoelectronic Applications

*Zhaojun Li[1,2], Hope Bretscher[1,3], Akshay Rao[1*]*

[1]Cavendish Laboratory, University of Cambridge, JJ Thomson Avenue, CB3 0HE, Cambridge, United Kingdom
E-mail: ar525@cam.ac.uk
[2]Molecular and Condensed Matter Physics, Department of Physics and Astronomy, Uppsala University, 75120 Uppsala, Sweden
[3]The Max Planck Institute for the Structure and Dynamics of Matter, 22761, Hamburg, Germany

**Abstract**

The interest in obtaining high-quality monolayer transition metal dichalcogenides (TMDs) for optoelectronic device applications has been growing dramatically. However, the prevalence of defects and unwanted doping in these materials remains a challenge, as they both limit optical properties and device performance. Surface chemical treatments of monolayer TMDs have been effective in improving their photoluminescence yield and charge transport properties. In this scenario, a systematic understanding of the underlying mechanism of chemical treatments will lead to a rational design of passivation strategies in future research, ultimately taking a step toward practical optoelectronic applications. We will therefore describe in this review the strategies, progress, mechanisms, and prospects of chemical treatments to passivate and improve the optoelectronic properties of TMDs.

# 1. Introduction

Since the successful exfoliation of monolayer graphene, which illustrated that two-dimensional (2D) material can exist stably in ambient conditions, research attention on various 2D materials has grown significantly.[1–5] Among 2D materials, transition metal dichalcogenides (TMDs), with the chemical structure MX$_2$ (M=Mo, W; X=S, Se, Te), have been of great promise for optoelectronic applications due to their important physical and optical properties such as direct bandgaps, strong light-matter interactions, superior mechanical flexibility, as well as chemical and thermal stability, to highlight a few.[6–12] Many proof-of-concept 2D-TMD-based optoelectronic devices with high-performance have been demonstrated.[9,13–19] However, in general, monolayer TMDs are imperfect and often contain various intrinsic and extrinsic defects.[20] Despite the remarkable potential of TMDs and the progress towards various optoelectronic applications, many challenges in improving their intrinsic qualities, like mitigating many-body effects (trions) and defects, persist. In contrast to bulk TMD materials, the excitons (electron-hole pairs) are strongly confined to the monolayer plane and also experience reduced screening due to the change in the dielectric environment.[21] The strong Coulomb interaction results in tightly bound excitons with binding energies of several hundred meV, which dominate the optical and charge-transport properties of 2D TMDs.[22–24] The quasiparticles, so-called trions, formed by excitons and induced free charges through defects or adsorbates, have also been identified even at room temperature due to the strong electrostatic interactions and exhibit binding energies of tens of meV.[25] The background carrier concentrations can therefore play important roles in the recombination pathways and affect the optical and electronic properties of 2D TMDs. Additionally, oxygen and other adsorbates in the ambient condition can alter the properties or degrade the qualities of 2D TMDs.[26,27] Substrate-supported TMD monolayers also suffer from strain variations introduced by the roughness of the substrates, which modifies their electronic bandgaps.[28] More importantly, the covalent bond strength within the TMD monolayers is weaker due to the reduced dimensionality, and thus the 2D TMDs are predisposed to form atomic defects, such as vacancies, self-interstitials, grain boundaries, *etc*.[20,29–32] The electrons and holes in TMDs can be trapped in defect-resulting potentials, leading to localized excitons and non-radiative recombination pathways, which strongly influences the optical and electronic properties of TMDs.[33–35]

To this end, extensive efforts have been devoted to exploring approaches for preparing and improving the quality of semiconducting 2D TMD materials. The fact that 2D materials are essentially all surfaces provides a unique opportunity for controlling and tuning their optical and electronic properties. 2D TMD layers are extremely sensitive to all influences of the surrounding environment, and their properties can therefore be easily modified by external variables. Substrate engineering like using a thin flake of hexagonal boron nitride (h-BN) as an interfacial layer reduces the structural damage and the associated interface states, which leads to intrinsic optical properties of 2D TMDs accordingly.[36–38] Strain engineering has also emerged as a powerful strategy for tuning the optical bandgaps of TMD materials.[39,40] Additionally, the surface chemical strategy, a versatile and non-destructive method, is one of the most effective approaches to tailor the properties of 2D TMD materials for practical device applications.[41] Compared to other approaches, chemical treatments are advantageous as they can be made compatible with other processing steps required for scaling-up device fabrication for commercialization. However, designing treatments compatible with industry-scale device processing will require a precise understanding of the mechanisms behind known chemical treatments. Thus, setting up rational selection rules for chemicals to increase the potential of 2D TMDs in practical optoelectronic applications is of crucial importance. Previous review articles have discussed tuning optoelectronic properties by strain or substrate engineering, and the observed effects of various chemical treatments.[41–44] However, to the best of our knowledge, no reviews addressed both the effects and mechanisms behind these effects of surface chemical treatments on 2D TMDs. To this end, this review focuses on the recent efforts using various surface chemical treatments to achieve high quality high-quality in regard to optoelectronic applications, paying particular focus on the mechanisms behind such treatments.

In this minireview, we start with an overview of the major studies from the past few years on chemical treatments which improve the semiconducting quality of 2D transition metal disulfides (TMDSs) including $MoS_2$ and $WS_2$ (Section 2). More specifically, Section 2.1 covers the characterization approaches of TMDs utilizing photoluminescence (PL) and electron mobility as the main quality indicators, as well as discrepancies reported among literatures, and Section 2.2 discusses the mechanisms of each chemical treatment including the conflicts among various research groups. Given that the understanding of how several chemical treatments work has evolved over the past few years, we think the discussion of discrepancies between different works is important to clarify the chemical selection rules and guide the further design of chemical treatment strategies for defect passivation and property control. In Section 3, we will then turn our focus to the major studies and accompanying mechanisms of chemical treatments improving the quality of $MoSe_2$ and $WSe_2$. In recent years, there have also been numerous studies on the preparation of high quality 2D $MoTe_2$, and it was theoretically predicted that Te vacancies can open a bandgap, which could be tuned by lattice strain or external force.[45,46] However, little research has been done up to this point on chemical treatments of $MoTe_2$, thus discussions of $MoTe_2$ and $WTe_2$ are not included in this minireview.[47] Finally, we present our concluding remarks including issues addressed by chemical treatments, challenges facing the chemical treatments strategy development, and an outlook of future research directions in this area.

## 2. Chemical treatments of $MoS_2$ and $WS_2$

*The intrinsic doping* of transition metal disulfides (TMDSs) may lead to the formation of positive or negative charged excitons (trions) that redshift and broaden the PL spectra. Control of the carrier density is effective to modulate the optical properties of monolayer TMDs induced by the many-body bound effect.

*Defects* attenuate properties and device performance. In light-harvesting devices, defects can be detrimental if they assist carrier recombination and reduce mean free paths of photoinduced carriers, consequently diminishing device performance. The predominant defect species as well as their percentages in 2D TMDs synthesized as monolayers with various methods or mechanically exfoliated (ME) from bulk crystals might be different, which would lead to different chemical treatment requirements. This might also be one of the factors that caused the discrepancy between the mechanisms disclosed in different studies. A large concentration of defects in ME $MoS_2$ monolayers observed in transmission electron microscopy (TEM) and scanning tunnelling microscopy (STM) were thought to be sulfur vacancies (SVs) which possess the lowest formation energy, and in contrast to atom dislocations observed at the grain boundaries in chemical vapor deposition (CVD) grown $MoS_2$.[48–50] Hong *et al.* reported that antisite defects with Mo replacing S atom are dominant point defects in physical vapour deposition (PVD) grown ML $MoS_2$, while SVs are predominant in ME and CVD-grown samples.[48] The defect density in the CVD-grown ML $WS_2$ interior is reported to be ~ 0.33 $nm^{-2}$ through atomically resolved scanning electron microscopy by Terrones and co-workers in 2017, which is four orders of magnitude higher than in mechanically exfoliated $WS_2$.[51] In addition, the density of sulfur vacancies near the edges is around three times higher than in the interior in the CVD-grown $WS_2$. Similar defect distribution in monolayer CVD-grown $MoS_2$ was reported by Schuck and co-workers.[52] On the other hand, Su *et al.* showed CVD-grown rhombic monolayer $MoS_2$ with PL intensity eight times stronger than CVD-grown triangular samples indicating low density of defects in rhombic monolayer $MoS_2$. This was attributed to SV passivation by oxygen atoms which is predicted through density functional theory (DFT) simulations to remove in-gap states.[53] Moreover, there is still debate on the origin of defects existing in TMDs. A previous study has shown that oxygen substitutions can be the dominant defect instead of sulfur vacancies and that differentiating between them is not possible using high-resolution TEM alone.[54] Therefore, we have carefully included synthesis methods of TMDs for different chemical treatments studies.

2.1 Characterization

2.1.1 Photoluminescence enhancement

*The PL intensity or photoluminescence quantum yield (PLQY)* serves as a quality indicator of 2D TMDs for optoelectronic applications as it is sensitive to the many body effects, defects and sub-bandgap states.[55,56] The structures of reported chemicals which led to PL enhancements were summarized in Fig. 1. Matsuda and co-workers reported the exciton PL enhancement of mechanically exfoliated (ME) $MoS_2$ by drop-casting p-type chemical dopants 2,3,5,6-tetrafluoro-7,7,8,8-tetracyanoquinodimethane (F4TCNQ) and 7,7,8,8-tetracyanoquinodinmethane (TCNQ).[57,58] Su *et al.* showed substantial PL intensity enhancement of ME monolayer $MoS_2$ through physisorption of $H_2O_2$ as a p-dopant.[59] Tongay *et al.* reported over 100 times improvement of PL intensity of ME monolayer $MoS_2$ by physical adsorption of electronegative molecules like $O_2$ and $H_2O$.[60] The authors also illustrated that the charge transfer from $MoS_2$ to $O_2$ reduced the original sheet carrier density of $MoS_2$ as much as 0.5 $nm^{-2}$ assuming that one $O_2$ molecule was physisorbed on each unit cell of $MoS_2$ *via* DFT calculations. Similar results were obtained by Peimyoo *et al.* through drop-casting F4TCNQ and $H_2O$ on ME monolayer $WS_2$, as well as by Nan *et al.* through high temperature annealing ME monolayer $MoS_2$ due to $O_2$ bonding.[61,62] As shown in Fig. 2, the reduction redox of these molecules lie below the conduction band minima (CBM) of $MoS_2$ and $WS_2$, so charges can be depleted from intrinsically n-doped $MoS_2$ and $WS_2$. The p-doping effect was evidenced by the blueshifted PL spectra, more positive threshold voltage of back gated TMD FETs, and red-shifted out-of-plane vibration ($A_{1g}$) peak in Raman spectra.[61,62] Sun *et al.* reported higher PL intensity in CVD-grown $MoS_2$ monolayers as compared to ME $MoS_2$ monolayers, which was attributed to the high p-doping effect of adsorbates in air.[63] Similarly, Xu *et al.* observed that the PL intensity of thermal vapor sulfurization (TVS) monolayer $MoS_2$ synthesized in vacuum was significantly attenuated relative to TVS monolayer $MoS_2$ synthesized in air, and proposed that the higher PL intensity of $MoS_2$ in air was facilitated by molecular adsorption ($O_2$, $N_2$, etc.) on SVs.[64]

In 2015, Javey and co-workers demonstrated near-unity PLQY with no change in the overall spectral shape for ME MoS$_2$ monolayers on oxide substrates through a chemical treatment by the nonoxidizing organic superacid bis-(trifluoromethane)sulfonimide (H-TFSI).[65] They also observed that the PL lifetime of MoS$_2$ was lengthened from roughly 250 ps to 10 ns after the H-TFSI treatment. Later on, Javey and co-workers reported an encapsulation approach with an amorphous fluoropolymer CYTOP and a subsequent H-TFSI treatment, which yielded near-unity PLQY in both ME MoS$_2$ and WS$_2$ monolayers with excellent stability against postprocessing.[66] They proposed that the strong protonating nature of the superacid removed the contaminants on the surface and suppressed defect-mediated nonradiative recombination in the monolayers. Goodman *et al.* reported that the deep trapped dark exciton states which were associated with native structural defects were responsible for the long PL lifetime of H-TFSI treated MoS$_2$, and the H-TFSI treatment reduced nonradiative recombination through these states.[67] The exact mechanism of H-TFSI treatment is, however, not fully understood, which has been investigated by a few research groups. This will be further discussed in detail in Section 2.2.

In 2017, Atallah *et al.* reported that charged defects in CVD-grown MoS$_2$ monolayers could be electrostatically passivated by ionic liquids (ILs) with a grounded metal contact, leading to up to two orders of magnitude increase in PL yield.[68] Similarly, Park *et al.* showed PL enhancement and defect passivation of CVD-grown ML MoS$_2$ with a ML of titanyl phthalocyanine (TiOPc).[69] Poly(3,4-ethylenedioxythiophene) polystyrene sulfonate (PEDOT:PSS) was reported to passivate the SVs in CVD-grown MoS$_2$ by a sulfur adatom cluster through a hydrogenation process confirmed by scanning transmission electron microscopy (STEM) images and X-ray photoelectron spectroscopy (XPS) measurements.[70] In that study, the electron concentration of MoS$_2$ after the treatment decreased by 643 times, and led to a work function increase of ~ 150 meV as well as enhanced PL intensity. Jin and co-workers demonstrated the passivation of SVs in both CVD-grown and ME MoS$_2$ monolayers *via* various thiol molecules, which led to enhanced PL intensity.[71] The authors used thiol molecules with F-containing ligands as markers, and the functionalized products were characterized with XPS and Fourier transform infrared (FTIR) spectroscopy. The reported mechanism of this chemical treatment is discussed in Section 2.2.2 and illustrated in Fig 2b. Yao *et al.* immersed CVD-grown WS$_2$ monolayer into sodium sulfide (Na$_2$S) solution and achieved enhanced PL emission with WO$_{3-x}$ defects passivation validated by XPS measurements.[72] In their study, the inhomogeneous PL emission in the inner and edge region of pristine WS$_2$ monolayer was attributed to the different charge populations and defect states across the monolayer area, which was clarified by the STEM images showing both SVs and W vacancies. The authors also observed redshift of the PL spectra of ML WS$_2$ after the Na$_2$S treatment, which was due to the increased formation of trions and biexcitons evidenced by steady-state low temperature and laser-power dependent PL measurements.[72]

2.1.2 Mobility improvement

*Mobility* serves as another quality indicator of 2D TMDs for optoelectronic applications, which is sensitive to charged impurities, traps, and structural defects both inside the material and at the dielectric interface due to their atomic thickness. Interface engineering like the use of crystalline h-BN and thiol-terminated SiO$_2$ substrates were found to effectively improve the device mobility by supressing the extrinsic scattering process and modifying the properties of TMDs.[73,74] N-doping like using hydrazine on the surface of MoS$_2$ flakes to increase the density of carriers also led to increased mobility of TMDs.[75] Radisavljevic *et al.* realized mobility of ~ 200 cm$^2$V$^{-1}$s$^{-1}$ in a FET with ME MoS$_2$ monolayer as a conductive channel and HfO$_2$ as a gate insulator.[76] In 2014, Yu *et al.* reported a high mobility > 80 cm$^2$V$^{-1}$s$^{-1}$ in backgated (3-mercaptopropyl)trimethoxysilane (MPS) treated ME MoS$_2$ FET at room temperature with SV passivation, revealing the potential of chemical treatments for achieving intrinsic charge transport of 2D TMDs.[77] The mechanism of MPS treatment is addressed in Section 2.2.2 together with other defect passivation approaches. In 2017, Neupane *et al.* demonstrated a carrier mobility increase of both ME and CVD-grown TMDS monolayers based FETs *via* methanol treatment.[78] They observed a concomitant enhancement in the PL spectral weights of trions, a redshift of Raman A$_{1g}$ mode, as well as upshifted peaks in the XPS spectra of TMD monolayers after the methanol treatment, which confirmed the n-doping effect. The authors also proposed that methanol contributed to the reduction of defects in TMD materials validated by the increased exciton absorption peaks and prolonged fluorescence lifetime of TMD monolayers after the methanol treatment.[78]

In 2019, Rao and co-workers reported greatly enhanced PL intensity of ME WS$_2$ monolayers *via* oleic acid (OA, shown in Fig. 1) treatment comparable to that of H-TFSI-treated monolayers, and simultaneously improved the mobility in WS$_2$-based FET devices due to defect passivation.[79] Recently, they reported a generalizable SV passivation protocol using a passivating agent (thiol, thiophen or sulfide, Fig. 1), followed by the H-TFSI treatment. This two-step chemical treatment simultaneously achieved improved mobility and an increase in PL intensity of both MoS$_2$ and WS$_2$ monolayers.[80] The detailed mechanism of this chemical treatments is discussed together with the H-TFSI treatment in Section 2.2.1.

2.1.3 Discrepancy

Even though defect passivation in 2D TMDSs was often correlated with PL enhancement as stated in Section 2.1.1, there were also chemical treatments reported which led to defect passivation without PL enhancement. Nguyen *et al.* investigated the effect of chemical treatments on the electronic structure of liquid phase exfoliated (LPE) MoS$_2$ nanosheets *via* a series of thiols. The studied chemicals were thiols with aromatic rings of different electron withdrawing capabilities as well as alkylthiols with different chain lengths. The authors observed redshifted PL spectra after the chemical treatments without significant changes of decay kinetics, and attributed these phenomena to the formation of shallow trap states upon functionalization through the defect sites of MoS$_2$.[81] In their report, the successful thiolation on the surface of MoS$_2$, resulting in cathodic valence and conduction band edge shifts of ~ 500 meV, was confirmed by both ATR-IR and XPS measurements. Pierucci *et al.* reported that the incorporation of atomic hydrogen in CVD-grown MoS$_2$ monolayer could saturate the SVs forming Mo-H bonds and preserve the well-defined electronic structure of MoS$_2$ monolayer evidenced by high resolution XPS measurements and DFT calculations.[82] However, they observed a decrease in PL intensity after hydrogenation which was explained by the suppression of PL originated by defects from MoS$_2$. The varied impact on PL intensity of TMDs caused by different defect passivation chemicals could also be ascribed to the opposite (n *Vs.* p) doping effect of the chemicals. Jung and co-workers reported the SV passivation on ME 4-layer MoS$_2$ with two thiol molecules: mercaptoethylamine with lone electron pairs served as an n-dopant leading to a decrease in PL intensity of MoS$_2$ after the treatment, while 1H,1H,2H,2H-perfluorodecanethiol caused a p-doping effect and resulted in an enhancement in PL intensity of MoS$_2$ after the treatment.[83] Moreover, Amsterdam *et al.* illustrated that the deposition of metallophthalocyanines (MPcs) on ME monolayer MoS$_2$ quenched the low-temperature defect PL, with the quenching efficiency decreasing in the order CoPc > CuPc > ZnPc.[84] The authors observed partial PL quenching of MoS$_2$ A-exciton peak after MPcs deposition, which was ascribed to the mutual charge transfer *via* the formation of a type II heterojunction.

## 2.2 Mechanisms of chemical treatments

### 2.2.1 Mechanisms of chemical treatments without defect passivation

*P-type doping* is the most common mechanism of chemical treatments presented since monolayer $MoS_2$ and $WS_2$ are intrinsically n-doped. Trions emit at longer wavelengths with an emission efficiency much lower than that of neutral excitons.[36] The charge transfer between the dopant and the 2D TMD material modulates the Fermi levels of the TMDs and results in the modification of optical and electronic properties of TMD monolayers.[85] For 2D TMDSs, P-doping promotes the emission of neutral excitons over trions, leading to an enhancement and blueshift in PL, while the defect states and basic electronic structures of the TMD material remain unaltered. The chemical structures of p-dopants reported and their electrochemical redox potentials, as well as calculated band alignment of 2D TMDs are summarized in Fig. 1 and 2, respectively. Zhang *et al.* investigated the doping effect on ME $MoS_2$, $WS_2$, $MoSe_2$, and $WSe_2$ monolayers with "Magic Blue" $[N(C_6H_4\text{-}p\text{-}Br)_3]SbCl_6$ as the p-dopant, and achieved PL enhancement for all four TMD materials.[86] The extent of doping level was modified by varying the concentration of dopant solutions and treatment time, and the authors confirmed the doping effect by transistor measurements, PL, Raman, and XPS spectroscopy. Birmingham *et al.* reported the effect of dopant phases (liquid or gaseous) on PL intensity of CVD-grown $MoS_2$ monolayer *via in-situ* Raman micro-spectroscopy, and concluded that the liquid dopant contributed lower charge transfer efficiency.[87] Wang *et al.* revealed that the effect of p-type doping on 2D TMDs not only depended on the chemical potential difference between the dopants and TMD materials, but also on the thermodynamic stability of physisorption by the means of temperature dependent PL measurements, gate-induced PL measurements and DFT calculations.[88] Rao and co-workers compared the effect of various chemical treatments including a series of ionic chemicals, H-TFSI, and small molecule p-dopants on the optical properties of monolayer TMDSs, and demonstrated that ionic salts like Li-TFSI, which are compatible with a range of green solvents, enhanced PL intensity of both ME $MoS_2$ and $WS_2$ monolayers to a level double that of H-TFSI treatment.[89] The authors revealed that both cations and counter anions play important roles in enhancing the PL intensity of TMDSs. The cations must be stably adsorbed on the TMDS surfaces, and the counter anions should be non-coordinating with strong electron-withdrawing groups. Their conclusions was supported by the appearance of the $A_{2u}$ mode in Raman spectra, cation adsorption *via* DFT simulation, time-resolved PL (TRPL) and PL diffusion measurements.[89] Recently, Zhou and co-workers demonstrated a universal p-type doping with Lewis Acid SnCl$_4$ *via* Sn$^{4+}$ ions exchange for TMDs, which is also proved by DFT calculation.[90]

*Cation intercalation* has been proposed as another mechanism of chemical treatments. The intercalated cations can result in p-type doping to the monolayer TMDs and reduce the substrate influence simultaneously. In 2017, Yu *et al.* demonstrated that the PL of CVD as-grown and transferred WS$_2$ monolayer was enhanced due to the intercalation of small cations (H$^+$ and Li$^+$) between the monolayers and underlying substrates, which was achieved by simply immersing substrate-supported monolayers into certain chemical solution.[91] The intercalation was evidenced by an increase in the atomic force microscopy (AFM)-measured height of the as-grown monolayers after the chemical treatment. Through a series of steady-state PL measurements, they also concluded that intercalation was less-likely to occur in TMD monolayers which interacted with substrates more strongly, like for as-grown monolayers or monolayers on 2D material substrates (h-BN, for example).

*The mechanism of H-TFSI treatment* has gained a specific focus since the chemical treatment has received the most attention in the past few years. In 2017, Kim and co-workers found that the H-TFSI treatment had a minimal effect on the inner region of triangular CVD-grown WS$_2$ monolayers, whereas the PL of WS$_2$ in the edge regions was enhanced up to 25 times.[92] They concluded that H-TFSI p-doped the sample, and reduced defects which they assumed were distributed unequally throughout the sample, thus leading to the spatially heterogeneous effects. Subsequently, Kim and co-workers reported that SVs in CVD-grown MoS$_2$ were directly repaired by the extrinsic sulfur atoms produced from the dissociation of H-TFSI, evidenced through a correlative combination of optical characterization, atomic-scale STEM and DFT calculations.[93] The detailed mechanism of the H-TFSI treatment proposed by the authors was shown in Fig. 3a, where H-TFSI molecule initially released SO$_2$, after which a SV was passivated, resulting in the dissociation of an O$_2$ molecule from the SO$_2$ after SV passivation. The authors also observed that the PL peak position of MoS$_2$ blueshifted by ~ 5 nm and the A$_{1g}$ Raman mode blueshifted by 0.6 cm$^{-1}$ after the H-TFSI treatment, suggesting a p-type doping effect. On the other hand, Kiriya *et al.* compared the effect of H-TFSI treatments in various solvents on the PL intensity of ME MoS$_2$ monolayer with H$_2$SO$_4$ and Li$_2$SO$_4$ in water, and concluded that the proton is a key factor to enhancing the PL intensity of MoS$_2$.[94] Relatedly, Lu *et al.* reported that the effectiveness of H-TFSI depended critically on the charge state and protons donated by H-TFSI.[95] According to their DFT simulations, three H atoms symmetrically adsorbed around the SV site in its -1 charge state, which could remove all gap states (Fig. 3b). Schwermann *et al.* revealed through first-principle calculations that H-TFSI transferred oxygen to the surface of monolayer $MoS_2$ yielding in a defect-free electronic band structure like that of pristine $MoS_2$.[96] The authors also pointed out that there were similar reactions with $H_2O_2$, $O_2$ and $H_2SO_4$ (but not $H_2O$) treatments, which was supported by simulations and steady-state PL measurements. Molas *et al.* observed that the H-TFSI treatment resulted in progressive quenching of defect-related emission in ME $MoS_2$ monolayers at low temperatures, again concluding the defect-passivating effect of H-TFSI treatment.[97]

Countering the growing body of claims that H-TFSI passivates defects through some mechanisms, in 2019, Javey and co-workers showed near-unity PL QY of pristine ME $MoS_2$ and $WS_2$ monolayers through electrostatic doping, and revealed that the underlying mechanism of the H-TFSI treatment is p-type doping without defect passivation further justified by the TRPL measurements, where the H-TFSI treatment led to similar decay kinetics compared to the electrostatic doping.[98] Their work implied that all neutral excitons in 2D TMDSs radiatively recombined even in the presence of native defects. On the other hand, Pain and co-workers reported PL enhancement of 2D TMDSs with superacid analogues and pointed out that acidity and the inclusion of sulfur and oxygen from H-TFSI did not necessarily play the roles in defect passivation.[99] Rao and co-workers observed much longer PL lifetime (1 ~ 20 ns) upon H-TFSI treatment compared to ME pristine $MoS_2$ monolayer which fell below the instrument response of 100 ps, indicating a trap-mediated exciton recombination process after H-TFSI treatment.[80] In addition, the authors experimentally observed sub-gap trap sites (originating from SVs) of $MoS_2$, which appeared as a positive feature at 730 nm in ultrafast pump-probe spectra. This sub-gap defect bleach grew simultaneously with the initial A exciton decay, confirming a transfer in population from the A excitons to the defect states. They also reported a decrease in carrier mobility by over two orders of magnitude in H-TFSI-treated FETs compared to untreated devices. The authors concluded that even though H-TFSI treatment increased the PLQY, the SVs were still present and significantly limited the quality of the TMD material. In addition, they have conducted a two-step chemical treatment, a passivating agent (thiol, thiophene or sulfide) followed by the H-TFSI treatment, which achieved enhanced PL intensity and shortened emission lifetime compared to H-TFSI-treated-only sample.[80] In contrast to the H-TFSI-only treatment, the sub-gap bleach was greatly reduced in the two-step treatment, suggesting the passivation of SV sites. The understanding of the mechanism behind H-TFSI treatment is then key to further design chemical treatments to passivate defects of 2D TMDs.

2.2.2 Mechanisms of chemical treatments with defect passivation

Defect passivation is defined as a process which removes the defect states from the energy gap between the valence and conduction band, without shifting the Fermi energy ($E_F$) into either band. At the time of writing this review, more investigations are required to identify the exact mechanism by which structural defects are repaired in TMD materials, as multiple proposals currently have been put forward in the literatures. A few groups reported that SVs in $MoS_2$ and $WS_2$ monolayers were chemically acting as catalytic sites for hydrodesulfurization reactions and, therefore, they could be passivated. This was proposed as the *sulfur vacancy self-healing (SVSH) mechanism*.[100,101] Yu *et al.* reported the reaction kinetics between the MPS molecule and $MoS_2$ simulated with DFT calculations.[77] As shown in Fig. 4a, the MPS molecule was chemically absorbed on the SV site of $MoS_2$ surface by cleaving the S-H bond. It formed a thiolate intermediate, and the dissociated H atom bonded to a neighbouring S atom. Then the S-C bond cleaved and formed the final product, trimethoxy(propyl)silane, after hydrogenation. The S-C bond in MPS was found to be weaker than other alkylthiol molecules due to the acidic nature of $CH_3$-O- groups and led to a low energy barrier for the reaction. In addition, the authors proposed that the $(CH_3O)_3$-Si- groups reacted with the $SiO_2$ substrate to form a self-assembled monolayer which passivated the $MoS_2/SiO_2$ interface. A similar mechanism was reported by Zhang *et al.* where the electrically neutral S adatoms filled the SVs of CVD-grown $MoS_2$ monolayer through a poly(4-styrenesulfonate) (PSS) induced hydrogenation process in mildly acidic PEDOT:PSS enviroment.[70,102] This finding was supported by both STEM and XPS measurements which showed that the contribution of the intrinsic $MoS_2$ species in the XPS spectra increased after the treatment.

The SV passivation with thiol chemistry has also been explored by multiple groups, yet both the resultant products and reaction mechanisms remain controversial. One reported mechanism of SV passivation is that thiol molecules conjugated to the TMD surface with the S-H bond cleaved rather than physisorption or chemisorption, as illustrated in Fig. 4b.[71,81,103–105] This was proposed as the *functionalization mechanism* and was often visualized by FTIR measurements in which the S-H band from thiol molecules was revealed at 2563 cm$^{-1}$, but absent after conjugation with $MoS_2$. Similarly, Cho *et al.* reported that alkanethiol molecules passivated the SVs through chemisorption at the SV sites of few-layer $MoS_2$, evidenced by a shift of the characteristic peak position in XPS after treatment.[106] On the other hand, McDonald and co-workers proposed that TMDs facilitated the oxidation of organic thiols to disulfides, which were physisorbed on the 2D TMD surfaces through electrostatic interactions, rather than coordinate at SVs.[107] This was proposed as the *dimerization mechanism*. The disulfide products were evidenced by the diffuse reflectance infrared Fourier transform (DRIFT) measurements. In this scenario, thiols initially donated a hydrogen atom to the TMD. The formed thiyl radicals yielded disulfides and the H[$MoS_2$] released hydrogen gas. The proposed mechanism is illustrated in Fig. 4c. Subsequently, McDonald and co-workers quantitatively monitored the consumption of 1-octanethiol in the presence of liquid exfoliated $MoS_2$ nanosheets using 1H nuclear magnetic resonance (NMR) spectroscopy and further confirmed the dimerization mechanism they proposed previously where $MoS_2$ facilitated the oxidation of thiols to disulfides.[108] In 2017, Wang and co-workers investigated the reaction mechanisms between defective $MoS_2$ monolayer and thiol molecules employing potential energy surface calculations and kinetic studies.[109] They concluded that the reactions were dominated by two competing mechanisms, dimerization or SVSH, and the dominant pathway was largely determined by the polarization of thiol molecules and the temperature. It is also worth noting that the authors predicted that two Mo-H species were yielded in the dimerization mechanism which is different from the other report about the same mechanism.[108,109] In 2021, Zhang *et al.* directly monitored the interaction between the fluorescent thiol and SVs in metal-organic chemical vapor deposition (MOCVD) grown $MoS_2$ monolayer *via* 2D point accumulation for imaging in nanoscale topography (PAINT) strategy, and revealed a hydroxide-assisted transition from the reversible interaction (physisorption) to covalent binding by deprotonation of the thiol while increasing pH.[110]

The SVs could also be passivated by the formation of a van der Waals interface, proposed as a *physisorption mechanism* like the use of ML TiOPc on $MoS_2$ surface.[69] Park *et al.* revealed a van der Waals interaction *via* scanning tunnelling microscopy (STM) and DFT modelling, in which negative charge transfer from $MoS_2$ to TiOPc removes defect states.[69] As illustrated in Fig. 4d, it was hypothesized that a thermally stable TiOPc ML was formed on $MoS_2$ surface, which did not induce physical reconstructions of defects. Similarly, Ahn *et al.* reported that MPcs passivated SVs in ME $MoS_2$, evidenced by the weakened PL peak at 1.79 eV (associated with excitons bound to defects) after the treatment in low-temperature PL measurements.[111,112]

Besides organic thiol molecules and MPcs, there were other approaches reported for passivating defects in 2D TMDSs. Tapasztó and co-workers reported that the defects in $MoS_2$ were resulted by $O_2$ oxidation, where $O_2$ spontaneously incorporated into the basal plane of monolayer $MoS_2$ during ambient exposure. The substitutional oxidation of $MoS_2$ could be fully recovered to pristine *via* annealing the oxygen-substituted $MoS_2$ in a $H_2S$ atmosphere at 200°C, which was evidenced by STM images and DFT calculations.[113] On the other hand, there were a few research groups that reported that the chemisorption of $O_2$ could passivate SVs of TMDSs and resulted in defect-free electronic band structure similar to that of perfect monolayer TMDSs.[62,96,114,115] Sivaram *et al.* proposed that $H_2O$ could passivate SVs in the CVD-grown $MoS_2$, but the reaction required photo-generated excitons to overcome a large absorption barrier (Fig. 3c).[116] The $H_2O$ molecule was physisorbed on the surface of $MoS_2$ by means of an empty antibonding orbital, then the exciton-mediated dissociation of $H_2O$ resulted in O atom bonded to the SV with a valency of -2 and $H_2$ molecule desorbed from the $MoS_2$ surface.

## 3. Chemical treatments of $MoSe_2$ and $WSe_2$

Up to date, there are few general chemical treatments enhancing the quality of both sulfur-based and selenide-based 2D TMD materials, which could be attributed to their different intrinsic doping and defects.[78,86] For instance, 2D $WSe_2$ is known to be a p-type semiconductor while $MoSe_2$ is known to be a n-type semiconductor due to their intrinsic defects.[117,118] In 2016, Javey and co-workers reported the effects of H-TFSI treatments on the PL QY of $MoS_2$, $WS_2$, $MoSe_2$, and $WSe_2$, and suggested that only the defects in sulfur-based 2D TMD materials were amenable to the H-TFSI treatment.[119] Later, they reported that ME $MoS_2$ and $WSe_2$ monolayers were hardly doped, and the electrostatic doping was, therefore, not able to enhance their emission.[98] This work bore some differences from a related study by Yu *et al.*, who reported a PL increase of the CVD as-grown $WSe_2$ and $MoSe_2$ monolayers after the H-TFSI treatment.[91] On the other hand, Ahn *et al.* reported that both n-type dopant zinc phthalocyanine (ZnPc) and p-type dopant zinc hexadecafluoro phthalocyanine ($F_{16}ZnPc$) resulted in quenching of PL from CVD-grown few-layer $MoSe_2$.[111] Recently, Rao and co-workers demonstrated that the OA treatment on ME $MoSe_2$ monolayers enhanced the PL yield of $MoSe_2$ by an average of 58-fold, while also improving spectral uniformity of brightness and reducing the emission linewidth.[120] The authors revealed trap-free neutral exciton movement in OA-treated $MoSe_2$ monolayers evidenced by steady-state excitation intensity dependent PL and TRPL studies, and thus postulated that the defect passivation scheme of the OA treatment was related to the selenium vacancy passivation through oleate coordination to Mo dangling bonds without distinguishable structural changes.

## 3.1 Mechanisms of Chemical Treatments

Han *et al.* reported that the HBr treatment enhanced the PL intensity of CVD-grown monolayer MoSe$_2$ more than 30 times through p-doping and defect-healing.[121] The p-doping effect was validated by the intensity increase and frequency upshift of A$_{1g}$ mode in Raman spectroscopy. Undesired oxidized Se$^{4+}$ and bridging Se$_2^{2-}$ defects were removed, which was visualized by the peak shift and full width at half-maximum (FWHM) decrease of Mo$^{(IV)}$ 3$d$, as well as the elemental ratio increase of the anion to Mo in XPS spectra. The possible mechanism of the HBr treatment proposed by the authors is illustrated in Fig. 5.

A few other molecules and treatment protocols have also been reported for passivating defects. Guo *et al.* proposed that the Se vacancies in MoSe$_2$ could be well passivated with halogen atoms (except F) by means of first principles calculation, through a p-doping process.[122] Mahjouri-Samani *et al.* reported that the vaporization of selenium in vacuum by a pulsed laser repaired Se vacancies in synthesized MoSe$_2$.[123] Lu *et al.* demonstrated passivation of Se vacancies by oxygen through a focused laser treatment in air on CVD-grown WSe$_2$, which was verified by enhancement in the PL intensity, improvement of the photoconductivity, increase of W oxidation ratio in XPS, as well as increase in thickness in AFM images.[124] Ahn *et al.* reported that metallophthalocyanines (MPcs) exhibited defect-healing effects on the surface of CVD-grown MoSe$_2$ monolayer, evidenced by the temperature-dependent blueshift of the band-gap, narrower PL bandwidth, and the suppression of mid-gap defect-induced absorption in the ultrafast-pump-probe spectroscopy.[111]

Moreover, Javey and co-workers reported a PLQY of ~ 60% in CVD-grown WSe$_2$ monolayers after undergoing a *solvent evaporation-mediated decoupling (SEMD)* process, which was also higher than that in ME WSe$_2$ monolayers by an order of magnitude.[125] They attributed the enhanced PLQY to reduced nonradiative recombination due to the release of built-in strain by decoupling the grown WSe$_2$ monolayer during the SEMD process, validated by electron diffraction, *in-situ* PL imaging, TEM and TRPL measurements. Similarly, Chen *et al.* demonstrated that the PLQYs of both MoS$_2$ and MoSe$_2$ were enhanced by the solvent with a moderate volatilization rate like ethanol.[126]

## Summary and Outlook

In this minireview, we provide an overview of the state-of-the-art chemical treatments and related mechanisms on TMDs. The low quality of 2D TMD materials has been a veritable bottleneck to the incorporation of 2D TMDs in practical optoelectronic devices. Thus, we focus more on strategies improving the intrinsic quality of TMDs by passivating atomic defects or reducing their inherent doping with surface chemical treatments and discuss the discrepancies of reported related mechanisms. The photoluminescence of monolayer TMDs and mobility of FET devices built on 2D TMDs are utilized as two most important quality factors to evaluate the effect of the chemical treatments. The PL and mobility of resulting FET devices have been hugely improved with varied surface chemical treatments in most cases. The mechanisms behind the chemical treatments have also been explored in detail both theoretically and experimentally for further development of chemical treatment approaches. We have thus tried to clearly present the research to date on chemical treatments and the mechanisms behind them in this minireview, and in some cases, the literature is converging on a unified picture. However, disagreement and discrepancies across different treatments and mechanisms remain. This must be resolved to move forward. As previously described, some of the discrepancies in disclosed mechanisms may be ascribed to the lack of comparison of various chemical treatments on TMD monolayers with the same synthesis method and under the same measurement condition, as well as to precision limit in characterization tools. Therefore, more attention should be paid to the synthesis of materials, and the experiments utilized for characterization should be carefully chosen and scrutinized to determine what can provide the most useful information toward determining mechanisms. For instance, the electrical transport measurements were utilized to support the defects passivation mechanism with enhanced mobility after chemical treatments, whereas the resulted mobilities were often limited by the contact between the monolayers and the electrodes, and chemicals like H-TFSI could corrode the contact, which led to an unfair comparison among various chemicals. From our perspective, an essential tool to characterize the effect of chemical treatments on 2D TMD materials is ultrafast spectroscopy, which can reveal the carrier dynamics associated with defects without the additional complications of contacts and electrodes. However, this should not be performed in isolation, as the variation of exciton dynamics associated with defects and doping renders multiple processes with a variety of time scales that can also overlap.[127] Advanced microscopy techniques such as cavity-enhanced extinction microscopy and single-molecule localization microscopy coupled with fluorescence labelling are reported to give new insights into the defects' properties of 2D materials.[110,128] Therefore, we contend that the most robust approach is to combine various experimental approaches to construct a hypothesis.

There are other challenges have to be addressed to take full advantage of 2D TMDs in practical optoelectronic applications. For instance, another area we identify for specific future work and possible commercialization is the improvement of liquid exfoliation (LE) of 2D TMDs.[129] Although the LE methodology and the quality of liquid exfoliated TMD nanosheets have been developed and improved constantly, the PLQY of liquid exfoliated TMD nanosheets remains an outstanding challenge, which limits the scalability of device application to a large extent.[43,130–133] Chemical treatments development towards being compatible with liquid exfoliation process could be extremely fruitful for possible applications in flexible optoelectronics.

## Conflicts of Interest

There are no conflicts to declare.

## Acknowledgements

A. R. is grateful to the invitation on this contribution from the editor office. This work is supported by . Z. L. acknowledges funding from the Swedish research council, Vetenskapsrådet 2018-06610.

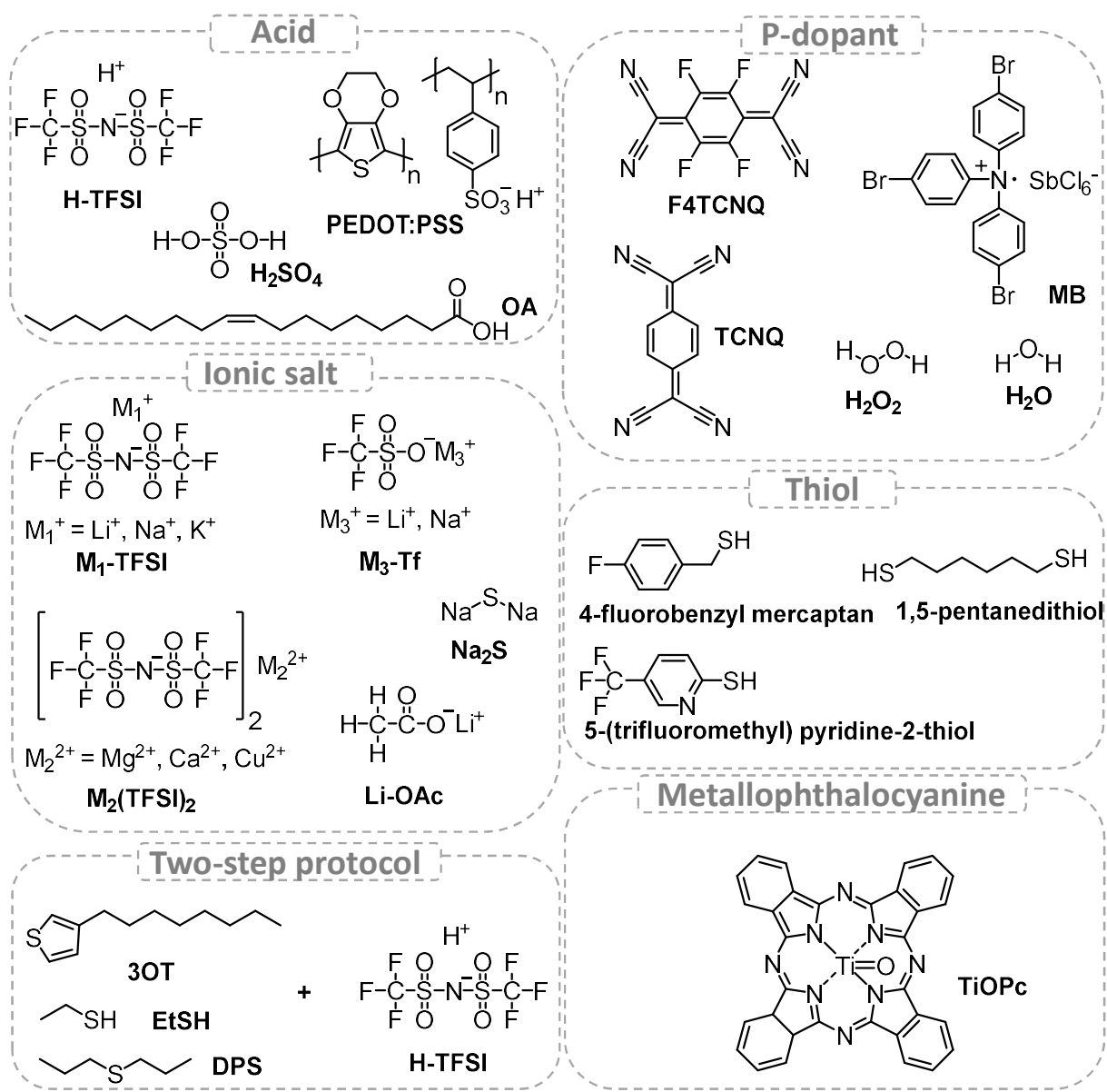

Fig. 1 Structures of chemicals used in previous studies for PL enhancement of 2D TMDs. Adapted from ref.80, 89.

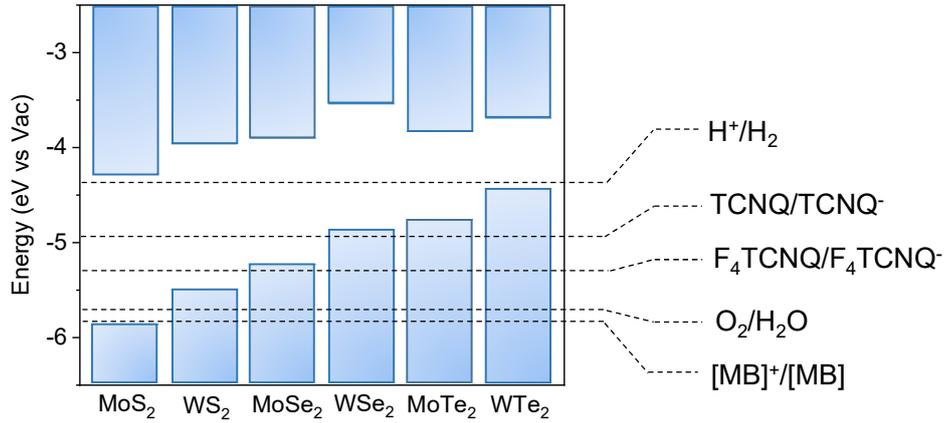

Fig. 2 Calculated band alignment of MoS$_2$, WS$_2$, MoSe$_2$, WSe$_2$, MoTe$_2$, and WTe$_2$ along with electrochemical redox potentials of the p-dopants from the literatures. Adliteratureref.41.

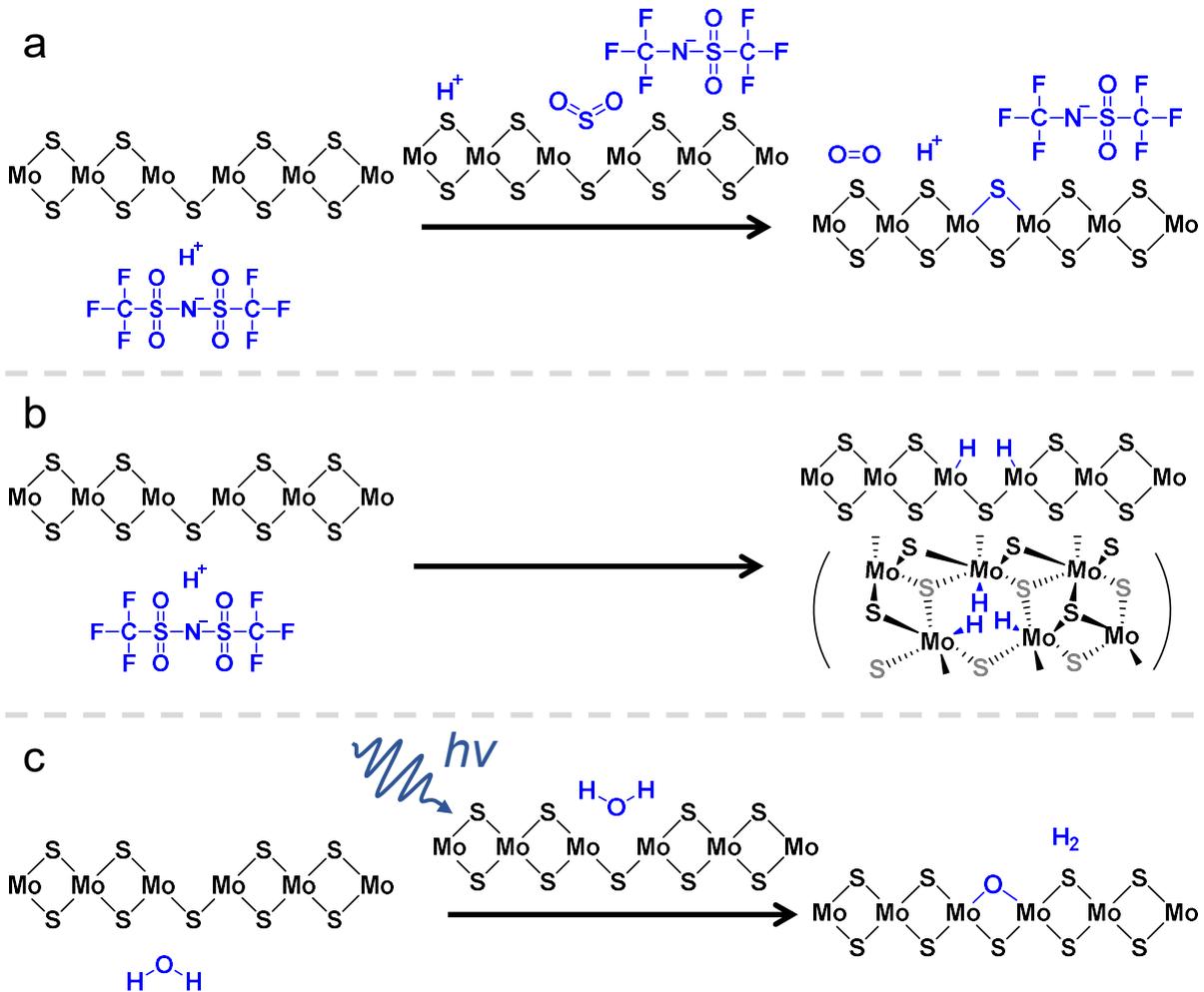

Fig. 3 Reported mechanisms behind H-TFSI and H$_2$O treatments for monolayer MoS$_2$ defect passivation. (a) illustration of the reaction between H-TFSI and monolayer MoS$_2$ through the SV sites. (b) illustration of the interaction between H-TFSI and monolayer MoS$_2$ through the

SV sites. (c) illustration of photon-mediated passivation with $H_2O$ molecule on the surface of $MoS_2$. adapted from ref.115.

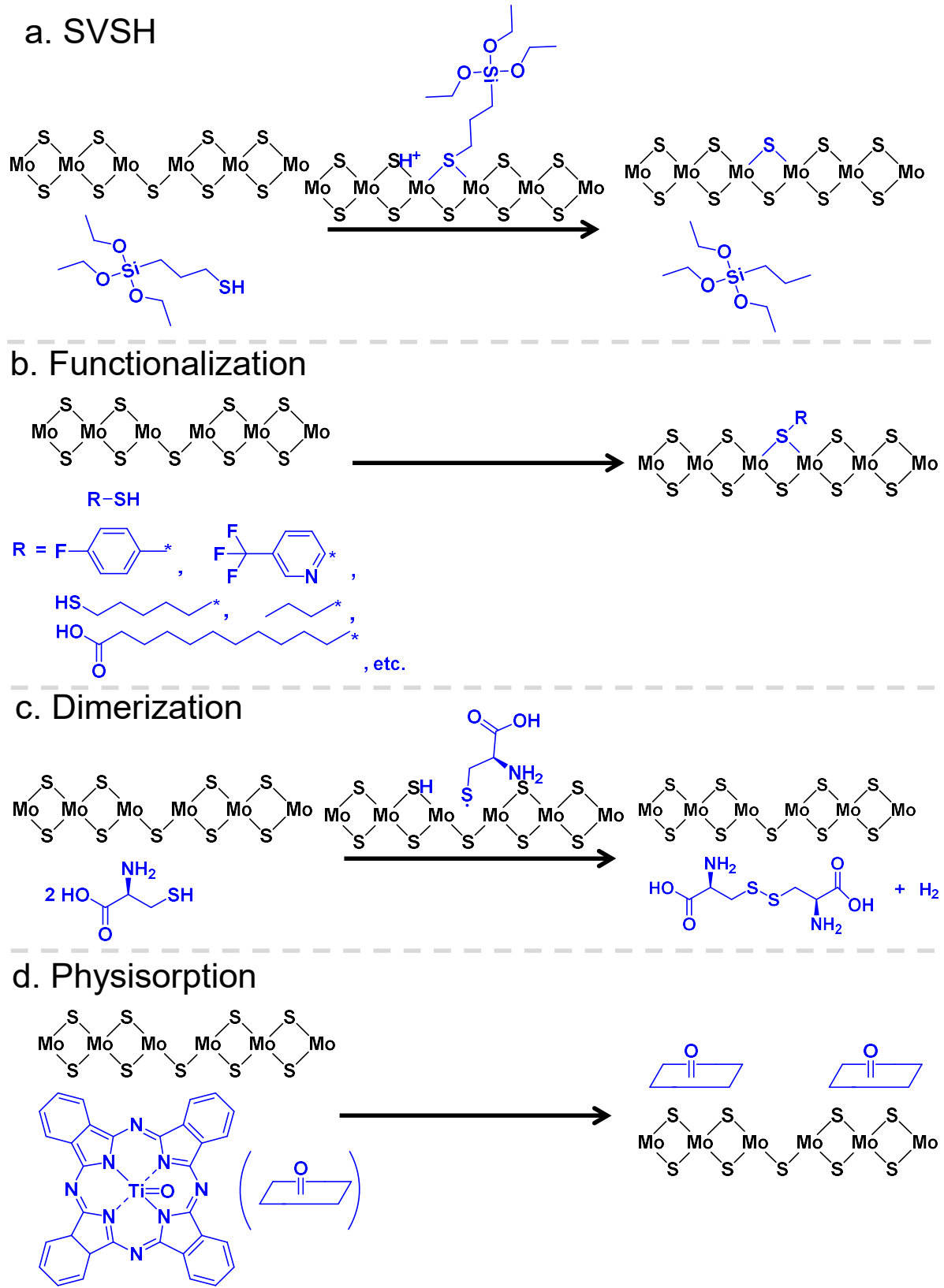

Fig. 4 Schematics of reaction kinetic the cs that have been reported for passivating the SVs. (a) the reaction between n MPS molecules and monolayer $MoS_2$ through the sulfur vacancy self-

healing (SVSH) mechanism. (b) the reaction between various thiol molecules and monolayer MoS$_2$ through the functionalization mechanism. (c) the reaction between the thiol molecule and monolayer MoS$_2$ through the dimerization mechanism. (d) interaction between TiOPc molecule and monolayer MoS$_2$ through physisorption mechanism.

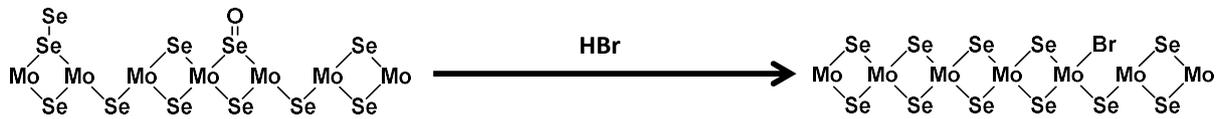

Fig. 5 Illustration of chemical reaction on the surface of MoSe$_2$ during the HBr treatment. adapted from ref.120.